# Two SK-I Neutrino Sources ?

Lasse E. Bergman
SIS, University of California, San Francisco
San Francisco, California 94143-0404

June 26, 2006

## Abstract

A search for a neutrino flux difference between weekdays and weekend days was undertaken for the average week of the Super-Kamiokande-I (SK-I) Experiment, using the 5-day period version of the SK-I data taken from May $31^{st}$, 1996 to July $15^{th}$, 2001. Arbitrarily uneven distributions of live time during the run time periods were considered before rejecting the null hypothesis. Such live time distributions were built into a robust method that calculated time-weighted neutrino flux means. The purpose was to show that the calculated results were unaffected by any distribution of live time, and thus that live time had no role in rejecting the null hypothesis. A significant ($p \ll 0.001$) difference was found and the most obvious neutrino flux change from weekdays to weekend days can be summarized as follows: "Some neutrinos took the weekend off – especially on Saturday".

## I. INTRODUCTION

This paper reveals a weekly periodicity of neutrino flux changes in the Super-Kamiokande (SK-I) data. However, the SK Collaboration [1,2,3] did not find any significant periodicity in the SK-I neutrino flux [4], thus ruling out "semiannual (seasonal) variations of the observed solar neutrino flux because of the changing magnetic field caused by the 7.25 degree inclination of solar axis with respect to the ecliptic plane" and any "short-time variation … due to the 27-day rotation of the Sun". So, with the exception of the Homestake experiment [7], "Kamiokande and other experiments have not provided any evidence for a time variation of the neutrino flux outside of statistical fluctuations [6] " [4].

The Collaboration applied the necessary correction for the uneven distribution of SK-I livetime during the 5-day runtime periods [4]. This crucial livetime factor has been overlooked (or by necessity had to be ignored) by other researchers [8] [9], who mistakenly claimed to have found a significant periodic variation of the SK-I data. The purpose of this paper is to demonstrate that a reported [5] neutrino flux difference between weekdays and weekend days was calculated by a robust method, i.e. that the calculated results were unaffected by any uneven distribution of SK-I livetime across the 168 (7*24=168) weekly hours.

## II. NO PERIODICITY

As the Collaboration members summarized their search findings, "We have presented the measured solar neutrino fluxes of … 5-day long samples using all 1,496 days [live time] of SK-I data. No significant periodicity was found in the SK-I solar neutrino data when a search was made to look for periodic modulations of the observed fluxes using the Lomb method. Based on a MC study, we have obtained the probability of finding a true periodicity in the SK-I data as a function of the modulation magnitude. The Lomb method should have found a periodic modulation in the SK-I solar neutrino data of … 5-day long samples if the modulation period were longer than … 20 days and its magnitude was larger than 10% of the average measured neutrino fluxes. …".

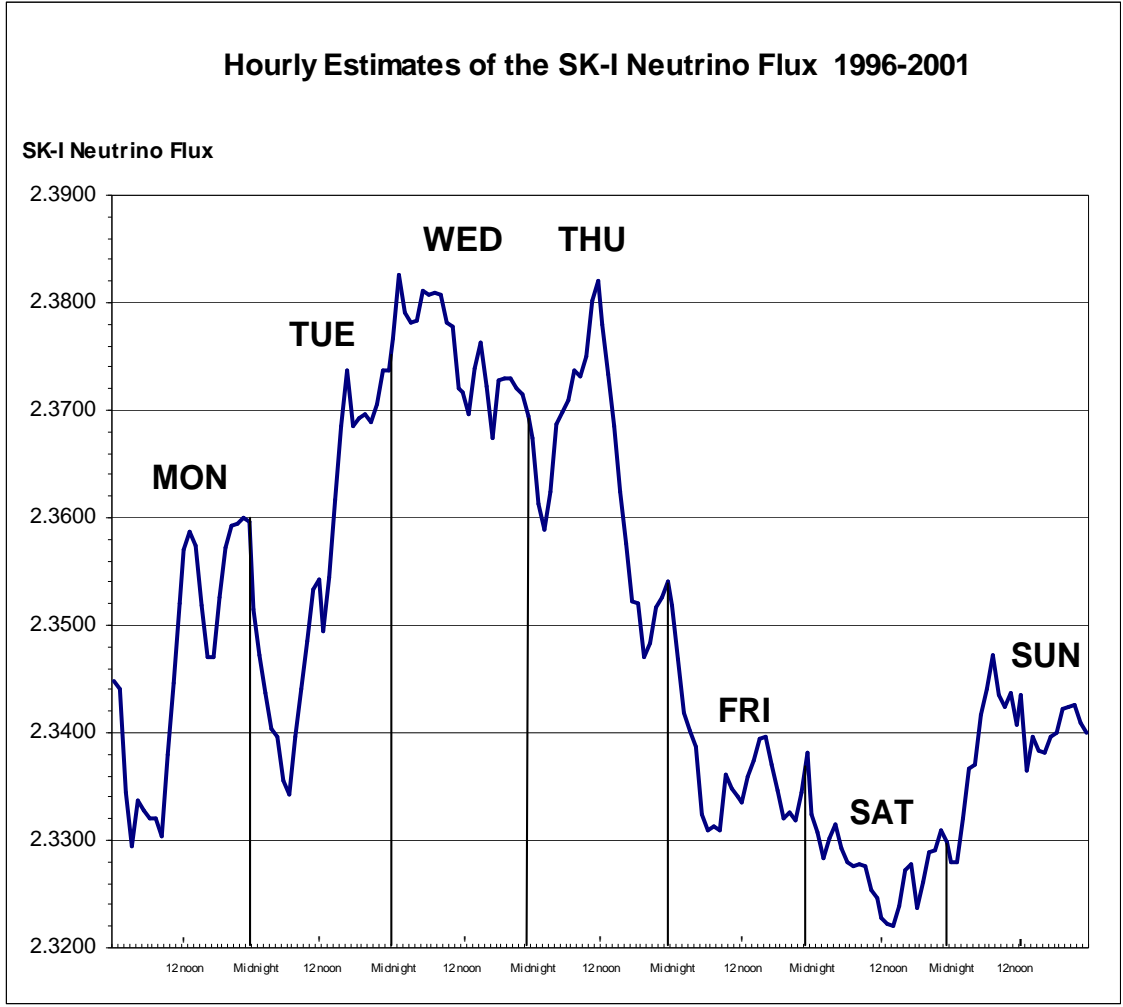

FIGURE 1: Hourly Estimates of the SK-I Neutrino Flux for the Average Week of 1996-2001. The horizontal axis is the day of the average week, and within each day are its 24 hours 0:00-0:59, 1:00-1:59, … , 23:00-23:59. The vertical axis is the weighted hourly estimate of the SK-I neutrino flux, for a particular day and hour. A weight (out of about (5/7)*358~260 weights for that particular day and hour) is the minutes (typically 60 minutes or less, depending on its livetime/runtime ratio) covered by that particular day and hour during a 5-day period in Table II in [4] and the neutrino flux value weighted by that weight is the SK-I neutrino flux for that particular 5-day period in Table II in [4]. The vertical axis has the units of $10^6$ cm$^{-2}$ s$^{-1}$. Please see Table I in [10] for more details.

## III.  TWO NEUTRINO SOURCES ?

Compared to the Collaboration's search for seasonal and short-time variations, this paper does find a pattern, but a much shorter, weekly one.  The period is shorter than 20 days, and its magnitude is smaller than 10%:  it is only 3% around the average measured flux of 2.33 x $10^6$ cm$^{-2}$ s$^{-1}$ [4] .   Finding such a modest weekly pattern is
3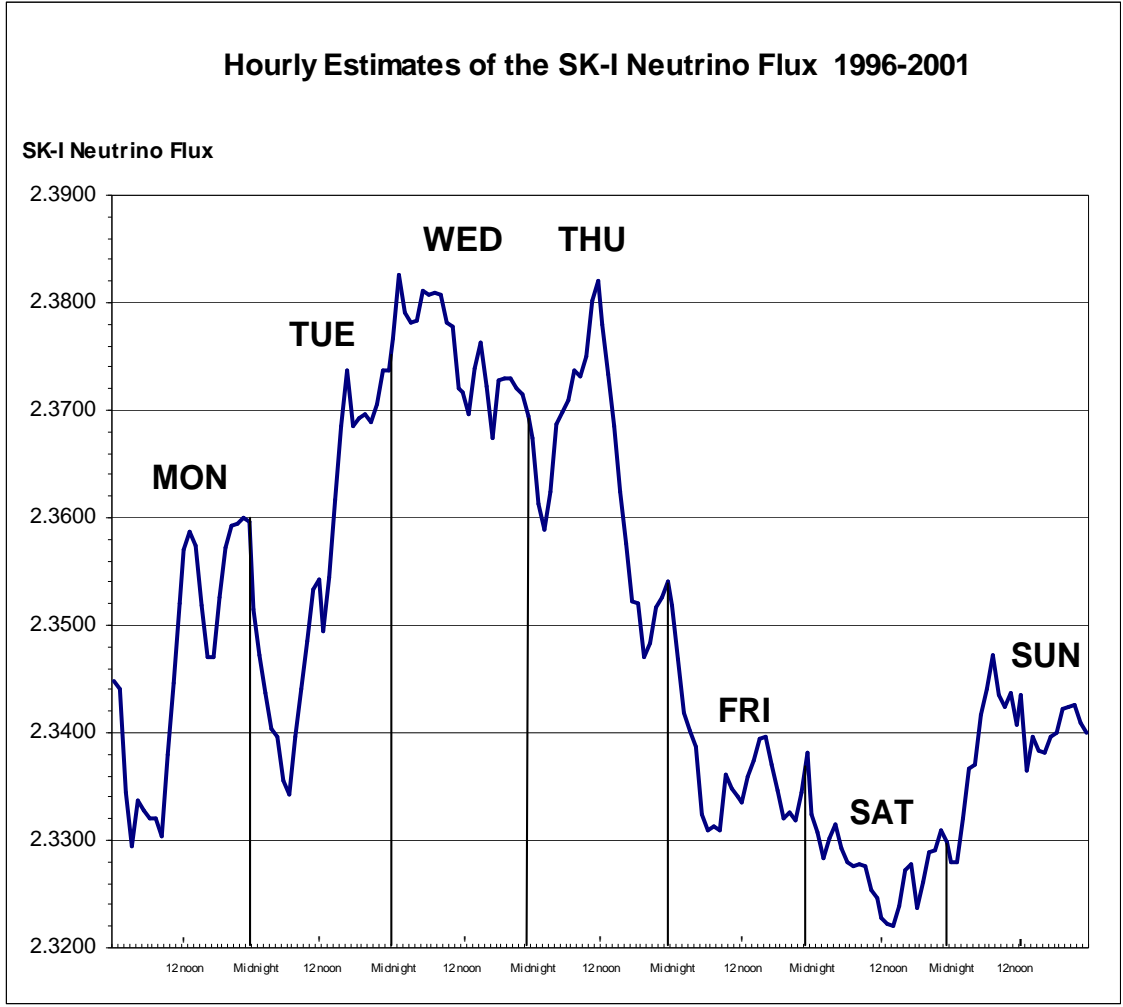

FIGURE 1: Hourly Estimates of the SK-I Neutrino Flux for the Average Week of 1996-2001. The horizontal axis is the day of the average week, and within each day are its 24 hours 0:00-0:59, 1:00-1:59, … , 23:00-23:59. The vertical axis is the weighted hourly estimate of the SK-I neutrino flux, for a particular day and hour. A weight (out of about (5/7)*358~260 weights for that particular day and hour) is the minutes (typically 60 minutes or less, depending on its livetime/runtime ratio) covered by that particular day and hour during a 5-day period in Table II in [4] and the neutrino flux value weighted by that weight is the SK-I neutrino flux for that particular 5-day period in Table II in [4]. The vertical axis has the units of $10^6$ cm$^{-2}$ s$^{-1}$. Please see Table I in [10] for more details.

## III.  TWO NEUTRINO SOURCES ?

Compared to the Collaboration's search for seasonal and short-time variations, this paper does find a pattern, but a much shorter, weekly one.  The period is shorter than 20 days, and its magnitude is smaller than 10%:  it is only 3% around the average measured flux of 2.33 x $10^6$ cm$^{-2}$ s$^{-1}$ [4] .   Finding such a modest weekly pattern is similar to spotting out needles in a hay stack, as the reader may notice by looking at Figure 4. More important, the shorter weekly period and its small magnitude can hardly be explained by solar neutrinos alone. A second neutrino emission source may have been present as well, registered by the same detector technology. In other words, the experimenters might have gotten "two neutrino sources for the price of one".

As in [10], the particular data set used in this paper was "collected at SK from May 31st, 1996 to July 15th, 2001, yielding a total detector live time of 1,496 days. This data taking period is known as SK-I", yielding some 15 events per day i.e. approximately 22,400 neutrino events for 1996-2001. The data set is the one arranged as 5-day periods, i.e. "neutrino data, acquired over 1,871 elapsed days from the beginning of data-taking, … divided into roughly … [5-day] long samples as listed in Table … " ( TABLE II in [4] ).

## IV. REJECTING THE NULL HYPOTHESIS

The results in this paper are visible in Figure 1, where each of the 168 "hourly estimates" from Table I in [10] has been plotted against its corresponding Day ( MON, TUE, WED, THU, FRI, SAT, or SUN ) and Hour ( "0:00-0:59", … , or "23:00-23:59" ). Figure 1 suggests that neutrino flux changes occurred from weekdays to weekend days. The most obvious change can be summarized as follows: "Some neutrinos took the weekend off – especially on Saturday". Figure 1 also displays local maxima at most (5 out of 7) midnights (cf. [5] ).

The 120 (5*24=120) weighted weekday means and the 48 (2*24=48) weighted weekend day means in Figure 1 and in Table I in [10] may also be viewed as two samples consisting of 120 and 48 plain values, respectively. Statistically, one can calculate the likelihood that these two samples were randomly drawn from identical populations, e.g. the likelihood (the p-value) that the difference between the two sample means was caused by chance alone. A (two-sided) two-sample t-test [9] for the difference between the two population means $\mu_{WeekdayHr}$ and $\mu_{WeekendHr}$ produced a significant
( $p = 7.32 * 10^{-16} \ll 0.001$ ) result, giving a strong reason to reject the null hypothesis
$H_o$: $\mu_{WeekdayHr} = \mu_{WeekendHr}$ .

The sample means were $\bar{y}_{WeekdayHr} = 2.3556$ and $\bar{y}_{WeekendHr} = 2.3335$, with the sample standard deviations $s_{WeekdayHr} = 0.016526$ and $s_{WeekendHr} = 0.007282$, and the sample sizes $n_{WeekdayHr} = 120$ and $n_{WeekendHr} = 48$. This significant result suggests that the neutrino flux was different between weekdays and weekend days.

## V.  CALCULATING A TIME-WEIGHTED MEAN

Figure 1 is thus based on the 168 neutrino flux estimates in Table I in [10]( 7 columns * 24 rows = 168 estimates) and each of those estimates is a weighted mean. Each term in the numerator of the weighted mean is a product of a neutrino flux value (the neutrino flux of a particular 5-day period) and a time weight (typically 60 minutes during that 5-day period). In Table I in [10] the number within parentheses, after the weighted mean, denotes how many products were used to calculate that particular mean.

For example, the first product (out of 264 products) for the column "FRI" and the row "4:00-4:59" in Table I in [10] has a time weight of 29 minutes and a neutrino flux value of 2.74 ( FRIDAY 05/31/1996 4.31 thru 4:59 in 5-day period  No. 1  in Table II in [4] where the neutrino flux for this first 5-day period is $2.74 * 10^6$ $cm^{-2}$ $s^{-1}$). To exemplify further, the second product ( out of 264 products) for the column "FRI" and the row "4:00-4:59" in Table I in [10] has a time weight of 60 minutes and a neutrino flux value of 2.83 ( FRIDAY 06/07/1996 4:00 thru 4:59 in 5-day period  No. 2  in Table II in [4] where the neutrino flux for this second 5-day period is $2.83 * 10^6$ $cm^{-2}$ $s^{-1}$). ). And, the third product ( out of 264 products) has a time weight of 60 minutes and a neutrino flux value of 2.30 ( FRIDAY 06/14/1996 4:00 thru 4:59).   And so on.  The "hourly estimate" for the column "FRI" and the row "4:00-4:59" in Table I in [10] was then calculated as the weighted mean  ( 29*2.74 + 60*2.83 + 60*2.30 + … ) / ( 29 + 60 + 60 + … ) = **2.3388** , in units of $10^6$ $cm^{-2}$ $s^{-1}$ .

## VI. RE-CALCULATING WITH DIFFERENT LIVE TIME

Please note that the "hourly estimate" above was calculated under the assumption that the live time was 100% of the run time <u>for the weekly hour "FRI 4:00-4:59"</u> (not necessarily 100% for others of the 7*24=168 weekly hours of which each could have a totally different livetime/runtime ratio). <u>This "hourly estimate" remains constant even under a different assumption regarding the livetime/runtime ratio</u> for "FRI 4:00-4:59". For instance, assume that the live time was 90% of the run time for "FRI 4:00-4:59". The time-weighted mean would then be re-calculated as ( 90%*29*2.74 + 90%*60*2.83 + 90%*60*2.30 + … ) / ( 90%*29 + 90%*60 + 90%*60 + … ) = **2.3388**, i.e. becomes the same calculated result as before, because all time weights were multiplied equally by the same 90% factor.

Figure 2 illustrates the SK-I run times distribution across the 7*24=168 weekly hours. As one may notice, the run times are higher at night and on Sunday. Not surprisingly, run time seems to be mimicked by down time: "… Now suppose for a moment that the detector has down time most often during the week, less often during the weekend. Suppose further on top of this that the detector has down time most often during the local day time, less often during the local evening time". [11]

Figure 3 simply illustrates that (not how) different live time distributions produce the same neutrino flux distribution, using the time-weighted mean method illustrated above.

## VII. A ROBUST CALCULATION METHOD

As demonstrated above, simple <u>separate</u> weighted average calculations (7*24=168 of them) would (for each of the 7*24=168 weekly hours) give a weighted average that is un-affected by a shifting live time distribution across the 7*24=168 weekly hours, because <u>all</u> the runtime weights for a certain weekly hour would be affected <u>equally</u> by the same livetime multiplication factor, i.e. the weighted average would remain the same. And the particular livetime multiplication factor would be exactly the livetime percentage that happened to be chosen for that particular weekly hour. Once an unchanging weighted average for each of the 7*24=168 weekly hours was established, one would then simply proceed to treat those averages as single numbers and apply more or less fancy techniques to describe them. As mentioned, in [10] a simple two-tailed t-test rejected the null hypothesis that the averages were the same for weekdays to weekend days.

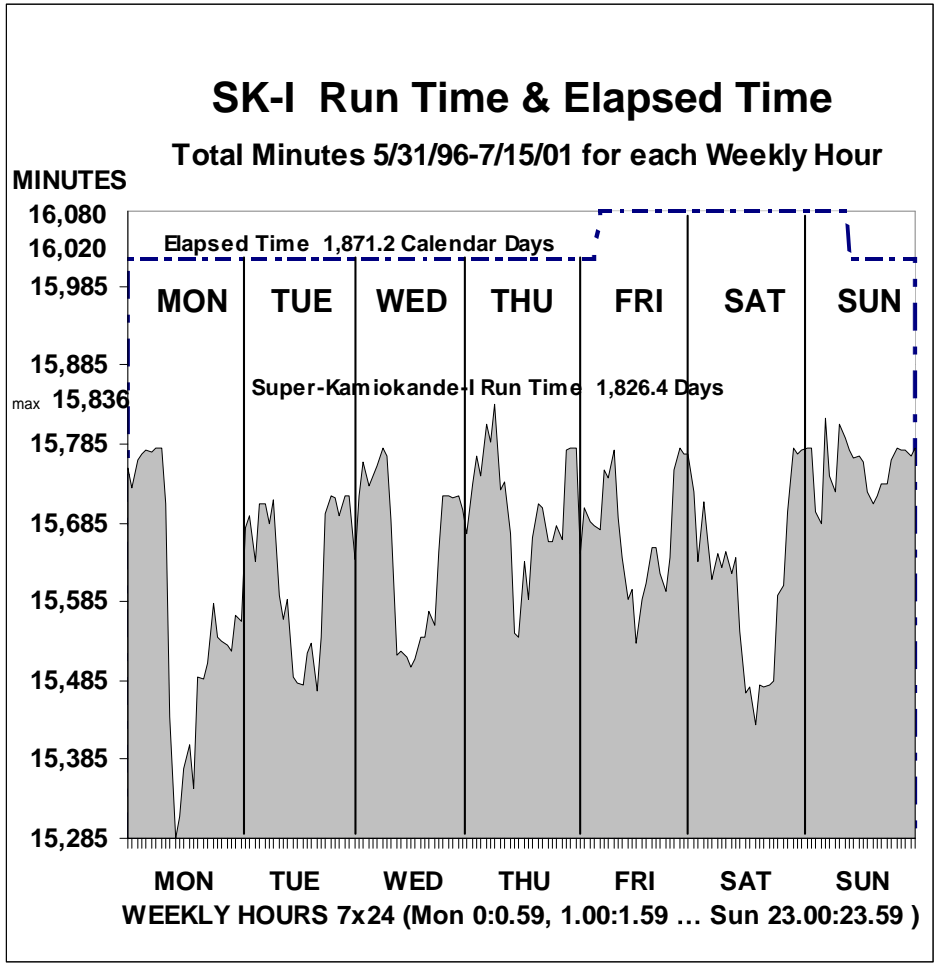

FIGURE 2: The SK-I Run Times distribution across the 7*24=168 weekly hours, for the Average Week of 1996-2001. The horizontal axis is the day of the average week, and within each day are its 24 hours 0:00-0:59, 1:00-1:59, … , 23:00-23:59. The vertical axis is Total Minutes during the period 5/31/96-7/15/01, for each of the 7*24=168 weekly hours. The vertical axis also shows Total Minutes of Elapsed Calendar Time for each such weekly hour. One may notice that run times are higher at night and on Sunday. Thus run time seems to be mimicked by down time: "… Now suppose for a moment that the detector has down time most often during the week, less often during the weekend. Suppose further on top of this that the detector has down time most often during the local day time, less often during the local evening time". [11]

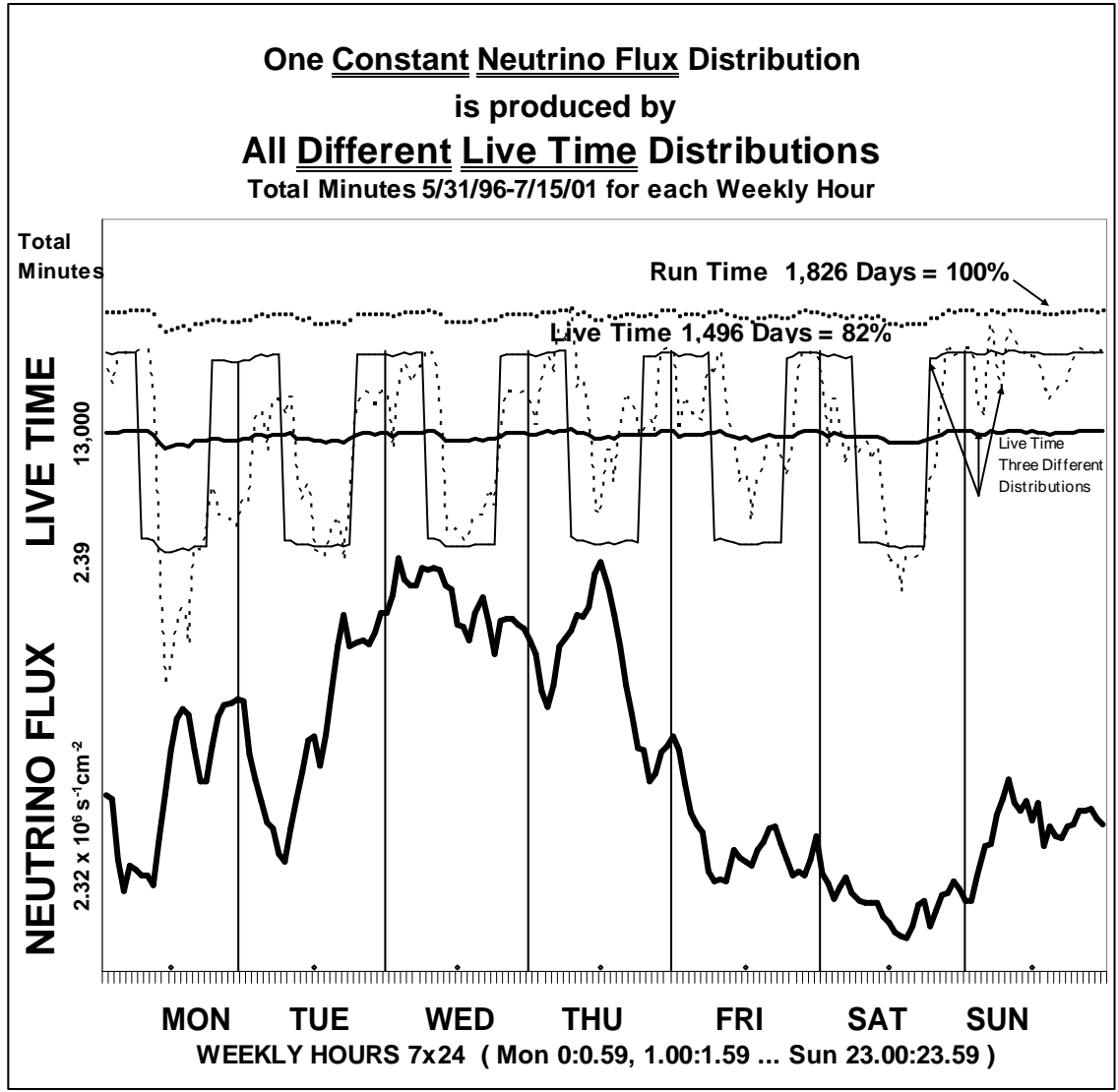

FIGURE 3: This figure simply illustrates that different Live Time distributions (each adding up to 82% of the total days of the Run Time distribution) produce the same Neutrino Flux distribution, when they are included in the method that calculates the time-weighted Neutrino Flux means. The horizontal axis is the day of the average week, and within each day are its 24 hours 0:00-0:59, 1:00-1:59, … , 23:00-23:59. The vertical axis serves two purposes: (a) denotes Total Minutes of Live Time during the period 5/31/96-7/15/01, for each of the 7*24=168 weekly hours; (b) denotes the time-weighted hourly estimate of the SK-I Neutrino Flux, for a particular weekly hour, identically to the vertical axis in Figure 1. The three Live Time distributions are just three (rather extreme) examples. One example mimics the Run Time distribution. The other two examples illustrate the assumption that "the detector was down … most often during the local day time, less often during the local evening time" [11].

## VIII. A NOT-SO-ROBUST CALCULATION METHOD

On the other hand, there are calculation methods that would be too sensitive to live time changes. For instance, an <u>in</u>advisable approach would be to use a one-way analysis of variance model (a simple linear model) applied <u>across</u> the 7 weekly days Mon-Sun (7 "treatments"). Such a model would test the null hypothesis that all 7 daily populations were equal, but would also require that certain calculations (e.g. between sum of squares calculations) would <u>mix</u> affected runtime weights from more than one day. And such <u>runtime weights for different days would be affected quite differently</u> by a shifting live time distribution across the 7 days.

## IX. LIVE TIME CONCERNS

The author has received [11] very helpful comments pointing out the role that livetime might play in wrongly rejecting the null hypothesis: " … Now suppose for the moment that the detector has down time most often during the week, less often during the weekend. Suppose further on top of this that the detector has down time most often during the local day time, less often during the local evening time. For obvious reasons you could imagine this is the case – people work during the day, and during the week-days, though in Japan Saturday is a work-day. Thus there is a very uneven distribution of down-time during the calendar week. … You know (we have published) the average live time. Assume most of the down time is during the week-days (just as a guess, make it 2/3 of the down time is scheduled during week-days, with the other 1/3 distributed randomly throughout the running period). Calculate a synthetic data set , and run said data set through your analysis. Do this many times. I suspect that you'll find that you can not reject the null hypothesis with your analysis methodology".

In fact, those comments from a collaboration member were the direct cause why the author explored if live time indeed was a false reason to reject the null hypothesis (" … I suspect that you'll find that you can not reject the null hypothesis with your analysis methodology."). As it turns out, the analysis methodology seems to have been robust enough, and the null hypothesis was rejected.

Figure 2 displays <u>run</u>time and <u>elapsed</u> calendar time, but not <u>live</u>time. One may already recognize, for runtime though, "that the detector has down time most often during the local day time, less often during the local evening time".

## X.   HOURLY ESTIMATES

By necessity, the neutrino flux numbers presented in this paper are estimates, and these estimates leave something to be desired, for the following reason.

Hourly comparisons based on raw data would require access to <u>un</u>-binned experimental data, e.g. the date and time for each individual neutrino event. However, there are good reasons why no event-by-event summary of the SK-I data is yet publicly available. As the author was graciously informed, the rationale for not releasing un-binned SK-I data publicly might roughly be expressed as follows: "At the event level, interpretation of systematic errors, calibrations, and background subtraction, gets quite complicated, such that for someone not close to the gory details of the detector and reconstruction software, doing a proper job gets very difficult. This seems to be a common problem in high energy physics, unlike astronomy where raw data is routinely made public." [12]. Consequently, the 358 <u>binned</u> neutrino flux values in Table II in [4] were published with statistical uncertainties (for example 2.74 +0.63 -0.53 $10^6$ cm$^{-2}$ s$^{-1}$, for the first 5-day period in Table II in [4] ). Fortunately, the random effects of this statistical variation of the published flux values tend to cancel out each other when some (5/7)*358~260 of those 358 values (Table I in [10]) are included in calculating the numerator of each weighted neutrino flux mean. Even better, those statistical uncertainties were found useful in order to run MC experiments. The results from 100 such experiments are presented in Figure 4 .

## XI.   RESULTS FROM MC EXPERIMENTS

As mentioned, TABLE II in [4] displays statistical uncertainties for the measured $^8$B neutrino flux, e.g. 2.74+0.63-0.53 $10^6$cm$^{-2}$s$^{-1}$ for the first 5-day period in TABLE II in [4]. Observing the 358 flux distributions in this table, one recognizes that those distributions seem to be skewed, i.e. that each of the 358 positive numbers have an absolute value larger than its negative counterpart.

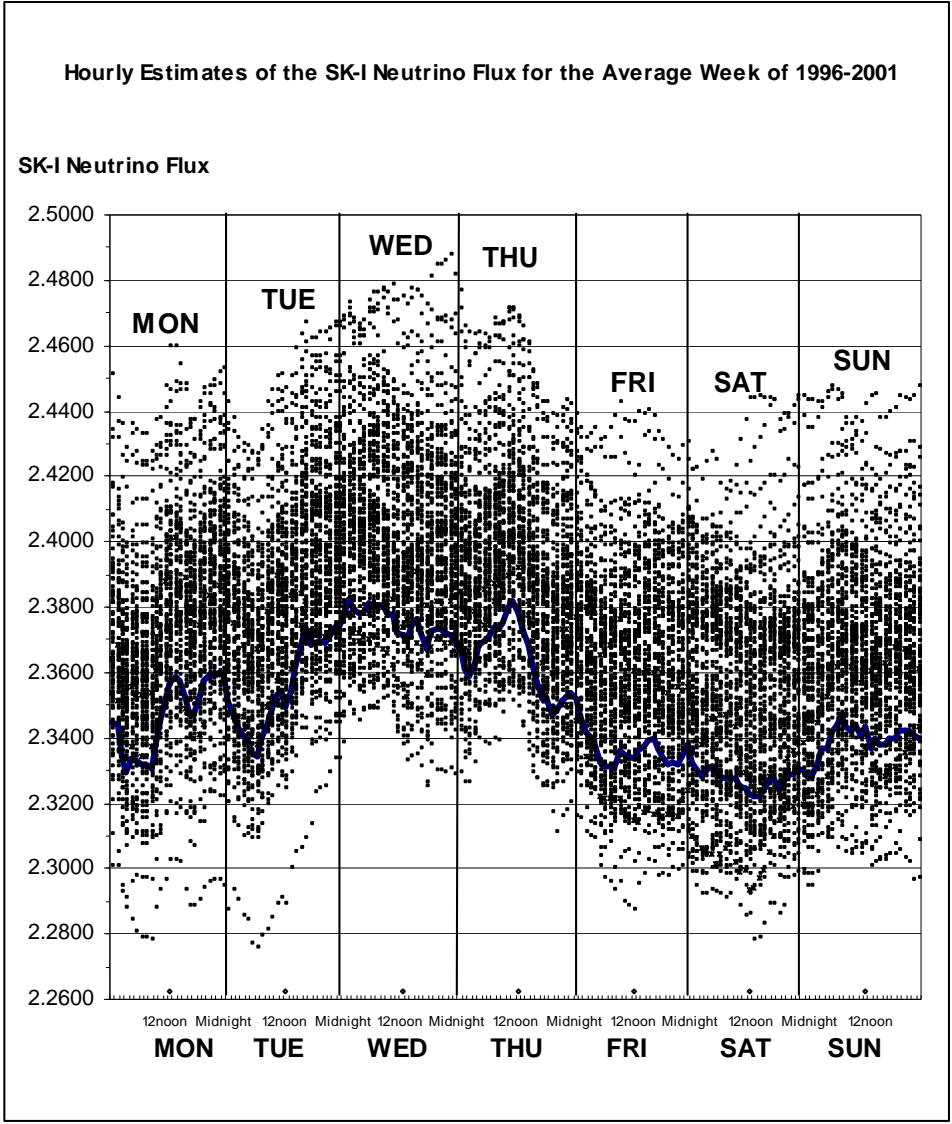

FIGURE 4: Hourly Estimates of the SK-I Neutrino Flux for the Average Week of 1996-2001. The thick flux line is the measured $^8$B neutrino flux in TABLE II in [4] as displayed in Figure 1. The 100 plotted lines, located asymmetrically around the thick flux line, are the results from the 100 MC experiments. The horizontal axis is the day of the average week, and within each day are its 24 hours 0:00-0:59, 1:00-1:59, … , 23:00-23:59. The vertical axis is the weighted hourly estimate of the SK-I neutrino flux, for a particular day and hour. A weight (out of about (5/7)*358~260 weights for that particular day and hour) is the minutes (typically 60 minutes or less, depending on its livetime/runtime ratio) covered by that particular day and hour during a 5-day period in Table II in [4] and the flux value weighted by that weight is the SK-I neutrino flux (measured or randomly generated) for that particular 5-day period in Table II in [4]. The vertical axis has the units of $10^6$ cm$^{-2}$ s$^{-1}$. Please see Figure 1 and the text for more details.

Taking a cue from a collaboration member ("Calculate a synthetic data set, and run said data set through your analysis. Do this many times."), the author used the statistical uncertainties of the flux values in TABLE II in [4] as parameters to run 100 MC experiments. The results from those simulations are presented in Figure 4.

The neutrino flux displayed in Figure 1 is included as the thick flux line in Figure 4. For each MC experiment, each of the 358 neutrino flux values in TABLE II in [4] was replaced by a flux value derived randomly as follows. As an example we refer to the first 5-day period in TABLE II in [4], for which the flux value is 2.74 . A random number between 0 and 1 was generated, e.g. 0.30 where the number was taken to define the left-most 30% of the area in a z-distribution (a normal distribution; yes quite an assumption), to the left of the flux value to be sought (30% cumulative normal probability). Since that random number was less than 0.50, the standard deviation |-0.53| was chosen, not |+0.63| in TABLE II in [4]. Next, that cumulative 30% area of the z-distribution corresponds to the z-value = -0.70. By adding the negative value -0.70 * |-0.53| = -0.37 to the first 5-day period's flux value 2.74, the random flux value 2.37 was produced. This random value was one of the 358 different (randomized) flux values used in one of the 100 MC experiments.

In Figure 4, the MC results show up as plots around the thick flux line, but in an asymmetric way because the 358 asymmetric flux distributions in TABLE II in [4] influenced the time-weighted mean method, asymmetrically, as follows. For each of the 100 MC experiments, roughly half (179 of 358) of the random flux values were smaller than their corresponding measured $^8$B neutrino flux values, and the other half of the random values were larger. But those larger random values were on the average farther from the $^8$B neutrino flux than those smaller random values. Therefore, the time-weighted mean calculation more often produced a weighted mean that was larger than the one in Figure 1, for the same weekly hour. The method thus less often produced a smaller one.

## XII.  DISCUSSION

For a possible reason why a neutrino flux difference (if any) between weekdays and weekend days should be expected, please see [5].

————————————————————

[1] Y Fukuda et al., Phys. Rev. Lett.  **81**,  1158  (1998).

[2] S. Fukuda et al., Phys. Rev. Lett.  **85**,  3999  (2000).

[3] Y. Ashie et al., Phys. Rev. Lett.  **93**,  101801  (2004).

[4] J. Yoo et al., Phys. Rev. D  **68**,  092002 (2003).

[5] L.E. Bergman, http://arxiv.org/abs/hep-ex/0504005  (2005).

[6] Y. Fukuda et al., Phys. Rev. Lett. **77,**  1683 (1996).

[7] R. Davis,  Prog. Part. Nucl. Phys. **32,**  13 (1994).

[8] A.Milsztajn,  http://arxiv.org/abs/hep-ph/0301252   (2003).

[9] D.O.Caldwell and P.A.Sturrock,  http://arxiv.org/abs/hep-ph/0305303   (2003).

[10] L.E. Bergman,  http://arxiv.org/abs/hep-ex/0509025   (2005).

[11] Personal communication from a SK-I Collaboration member (2005).

[12] Personal communication from another SK-I Collaboration member (2005).